\documentclass{article}

\usepackage{graphicx}
\usepackage[cmex10]{amsmath}
\usepackage{amsfonts}
\usepackage{subfigure}
\usepackage{url}

\begin{document}

\title{They Know Where You Live!}
\author{Kazem Jahanbakhsh, Valerie King, Gholamali C. Shoja\\
  Computer Science Department\\
  University of Victoria\\
  British Columbia, Canada\\
  \texttt{jahan@cs.uvic.ca}}
  
\maketitle

\begin{abstract}
In this paper, we demonstrate the possibility of predicting people's hometowns by using their geotagged photos posted on Flickr website. We employ Kruskal's algorithm to cluster photos taken by a user and predict the user's hometown. Our results prove that using social profiles of photographers allows researchers to predict the locations of their taken photos with higher accuracies. This in return can improve the previous methods which were purely based on visual features of photos \cite{Hays:im2gps}.
\end{abstract}

\section{Problem Definition: estimating place of living}
Suppose we have information about the places where a user (e.g. $u$) has visited during a period of time $T$. Let us denote every visited location like $l_{i}$ with a triple $(x_{i},y_{i},t_{i})$ where $x_{i}$ and $y_{i}$ show the latitude and longitude of the visited place and $t_{i}$ shows the visited time. Our problem is to predict the user $u$'s hometown given a sequence of visited locations with size of $n$. For the real data, we are going to use the available information from Flickr website \footnote[1]{www.flcikr.com}. In Flickr, users have the capability to upload and share their photos with other people in the community on the website. Users can also have a profile page where they can post their personal information such as their names, hometowns, gender, occupation, list of their friends, and so on. One interesting feature of Flickr website is that it allows users to add tags to each uploaded photo. More interestingly, users can geotag their photos by explicitly posting the location that they have taken a specific photo. In this paper, we assume that users only upload the photos which have taken by themselves. Thus, the geotag information of photos can be used as a proxy for user's mobility.

\section{Related Work}
Hayes et al. have already studied Flickr photos from an image processing point of view. They have proposed an image recognition algorithm which tries to predict the location of a photo by looking at the photo's visual feature \cite{Hays:im2gps}. Kalogerakis et al. have enhanced Hayes work by not only considering visual features of photos but also by taking into account the truncated Levy flight models of human mobility \cite{Kalogerakis:images2gps}. Kleinberg \cite{kleinsmall} and Nowell \cite{nowell} have independently shown that the probability of friendship for a given pair of users such as $(u,v)$ drops as the geographical distance between them (e.g. $d(u,v)$) increases. Motivated by their work, Backstrom et al. have proposed an algorithm for predicting a person's hometown by only having information about their friends' hometowns \cite{Backstrom:facebook}.

\section{Motivation}
To best of our knowledge, there is not any work in mobile computing area which has used Flickr's data to study human mobility models. In this work, we are going to show the possibility of using the Flickr data to study human mobility patterns. Estimating a user's place of living by studying their uploaded contents is an interesting problem which can have applications in social computing area. Furthermore, our work proves the effectivity of using people's social information for predicting the location of a given photo. We strongly believe that this method can enhance the previous algorithms which were based on visual features' of a single or a set of photos.

\section{How far do people take their photos from their hometowns?}
\begin{figure}
     \centering
     \includegraphics[width=4.00in, height=3.0in]{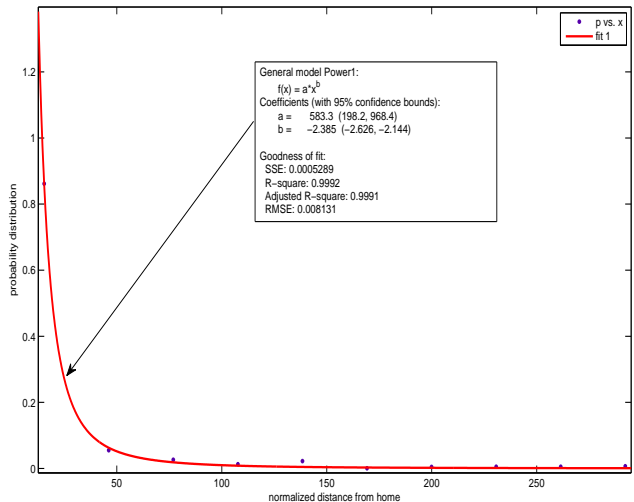}
     \caption{Distance probability distribution}
     \label{fig1}
 \end{figure}

To collect the Flickr photos and users' social profiles, we have developed several python scripts which use Flickr API \footnote[2]{http://www.flickr.com/services/api/} to crawl the requested data from Flickr website. In this paper, we take a random user (e.g. $u$) and collect all geotagged photos which have been uploaded by $u$'s friends. We also collect the hometown information for every $u$'s friend. Our chosen user $u$ has $31$ friends who have geotagged photos and have also reported their places of living on their profile pages. These $31$ people have totally uploaded $21219$ geotagged photos. First, we have computed the distance of each uploaded photo from its photographer's place of living. We have shown the probability distribution of photos' locations from their photographers' hometown in Figure \ref{fig1}. The distribution follows a power-law with exponent of $b=-2.38$ which is similar to proposed human mobility model by Gonzalez's which was obtained by analyzing cellphones data \cite{Gonzalez:nature}. This power-law distribution shows that people are more likely to take photos from places which are close to their home city as we expect intuitively. Even if this observation may seem natural, it gives us an effective way to predict the place where people live by analyzing their uploaded contents in virtual world.

\section{Hometown Predictor Algorithm}
\begin{figure}
\centering
\mbox{
\subfigure[osiatynska]{
\setlength\fboxsep{0pt}
\setlength\fboxrule{0.5pt}
\includegraphics[width=2.2in, height=1.8in]{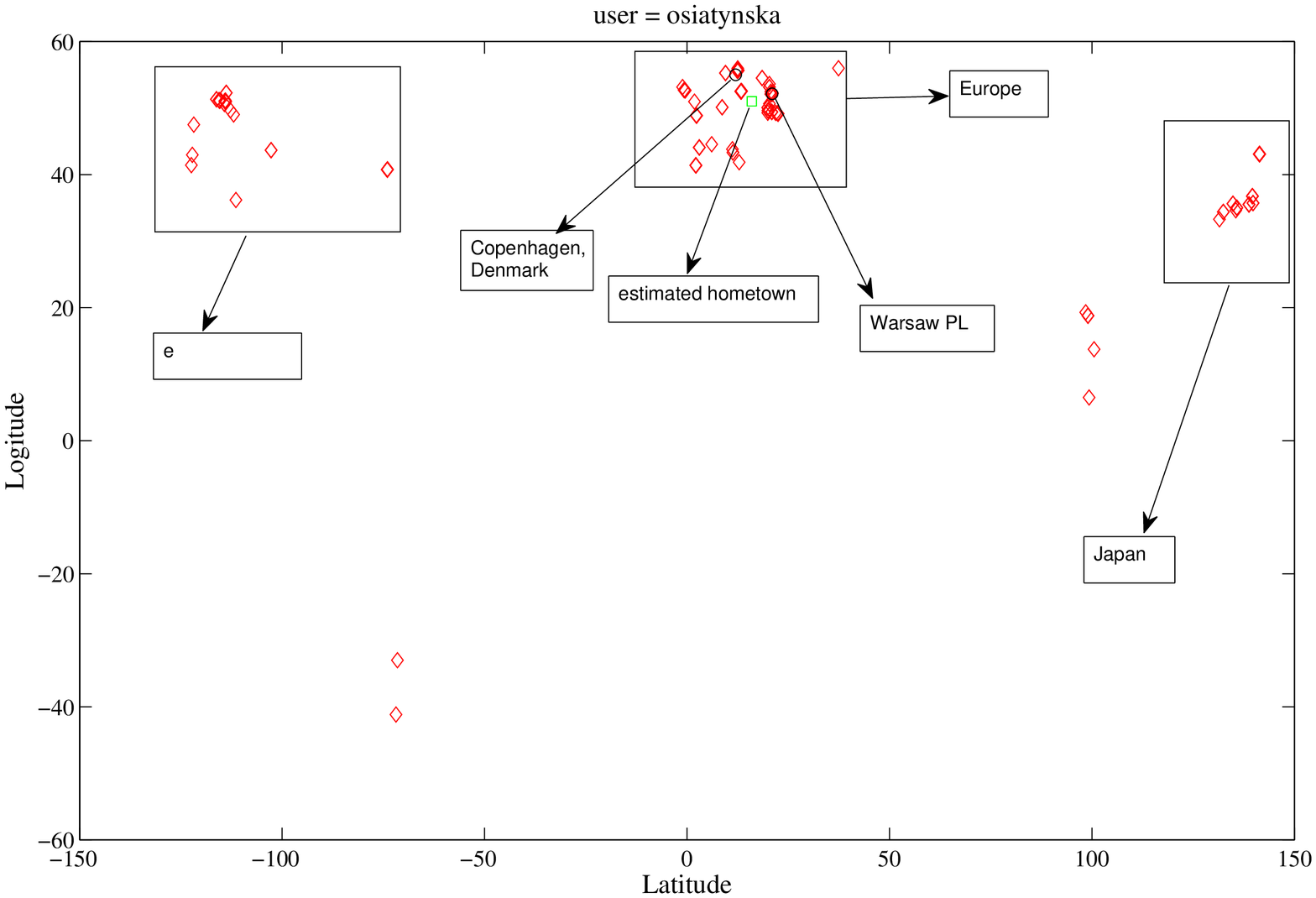}
\label{fig2:subfig1}
}
\quad
\subfigure[craignos]{
\setlength\fboxsep{0pt}
\setlength\fboxrule{0.5pt}
\includegraphics[width=2.2in, height=1.8in]{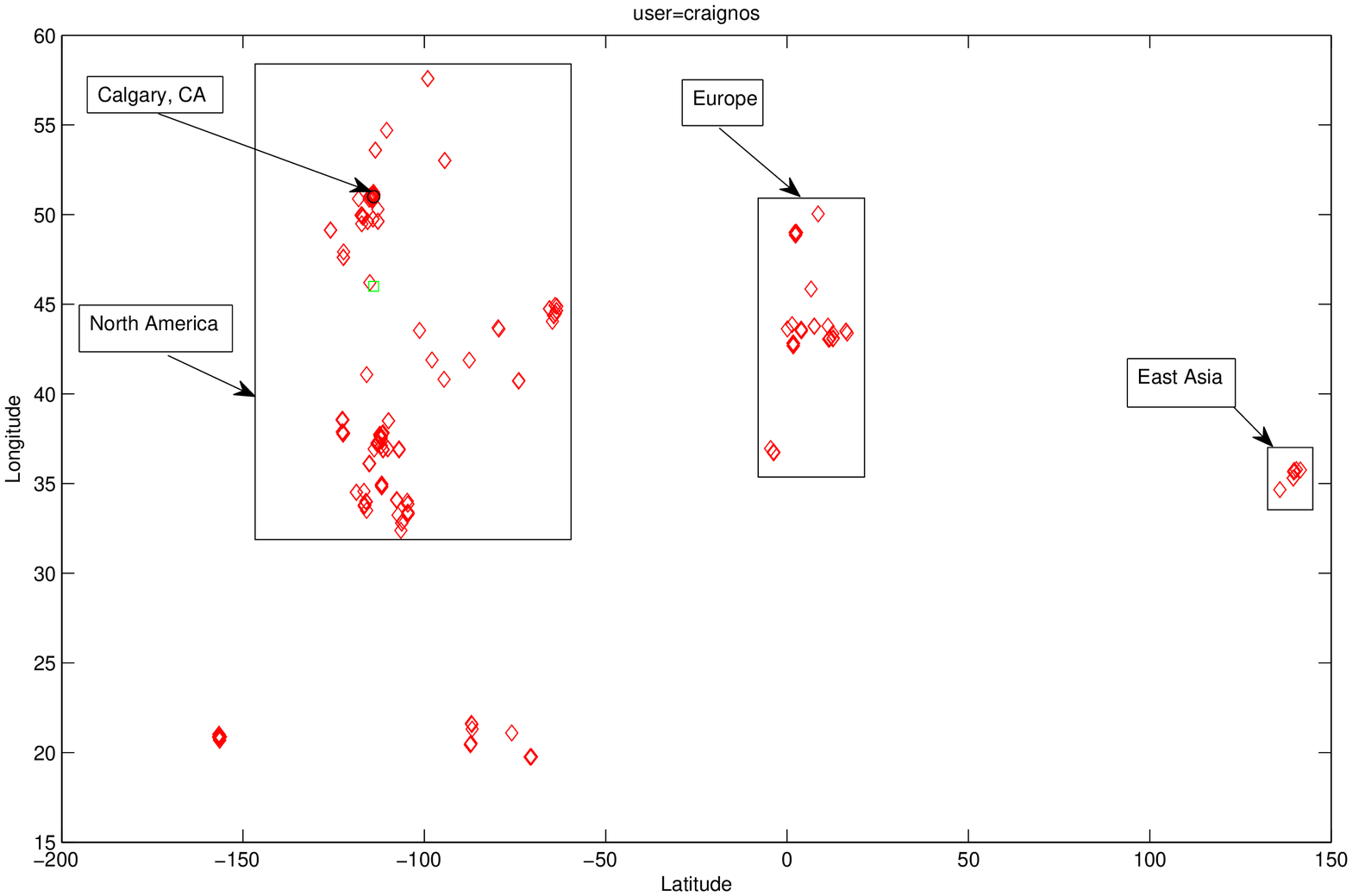}
\label{fig2:subfig2}
}
}
\caption{Photos' locations taken by two different users} 
\label{fig2}
\end{figure}

\begin{figure}
\centering
\mbox{
\subfigure[crosslens]{
\setlength\fboxsep{0pt}
\setlength\fboxrule{0.5pt}
\includegraphics[width=2.2in, height=1.8in]{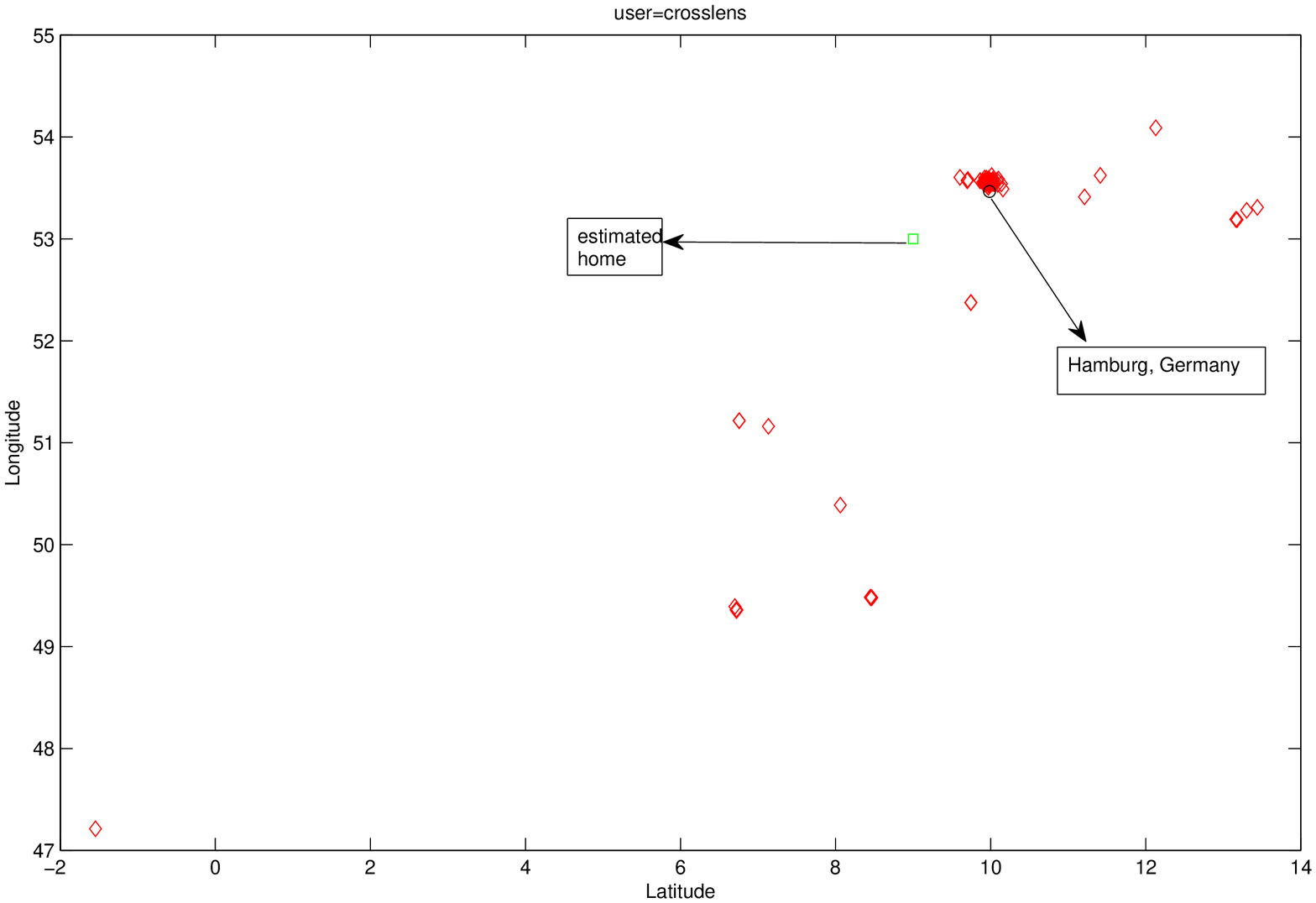}
\label{fig3:subfig1}
}
\quad
\subfigure[koltregaskes]{
\setlength\fboxsep{0pt}
\setlength\fboxrule{0.5pt}
\includegraphics[width=2.2in, height=1.8in]{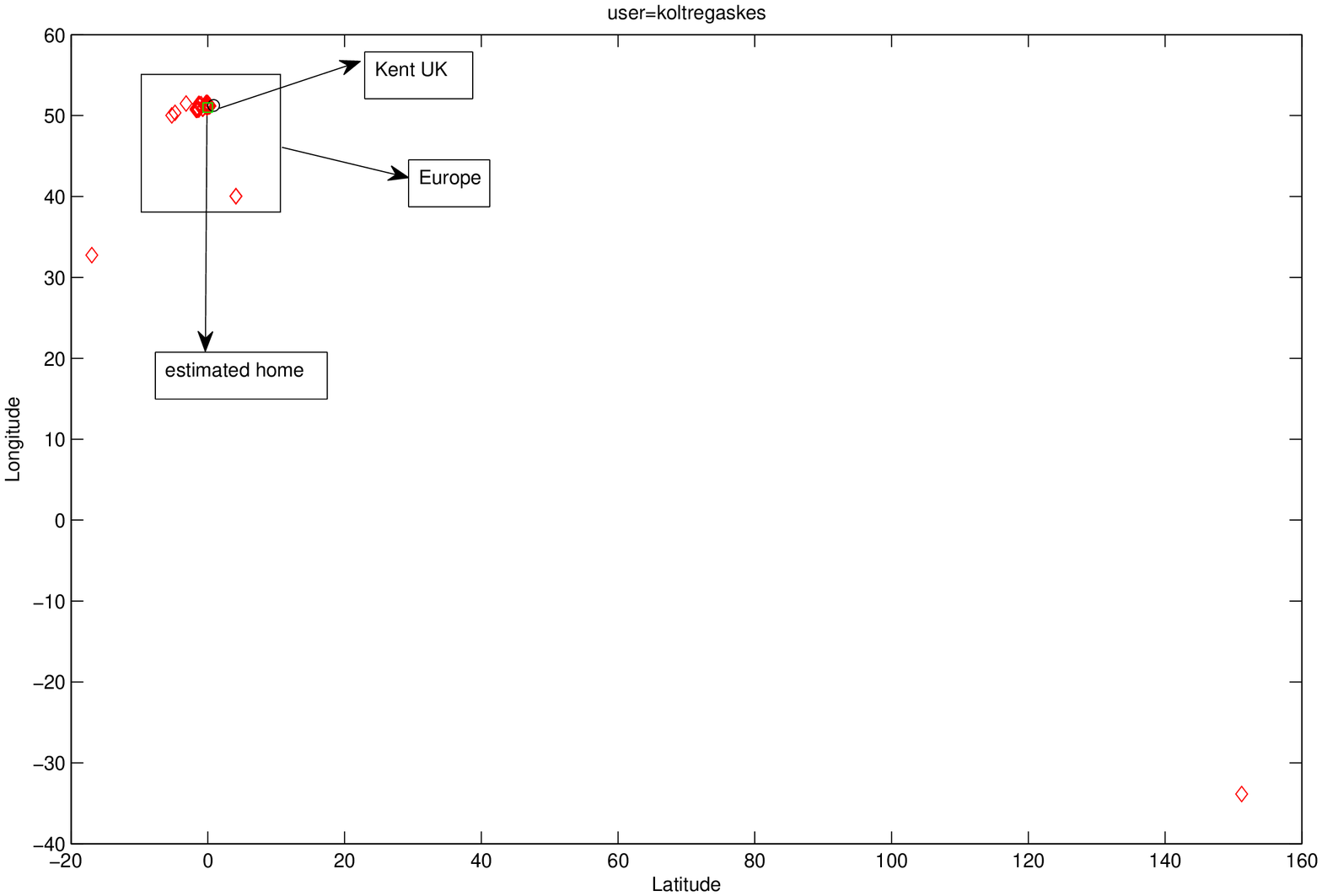}
\label{fig3:subfig2}
}
}
\caption{Photos' locations taken by two different users} 
\label{fig3}
\end{figure}

Considering our observation from the previous part, we can propose a simple but effective method to predict people's hometowns by analyzing their uploaded photos. Although people are more likely to take photos from places close to their home, they also take photos when they travel to other places far from their home cities or countries. Therefore, we need a clustering approach to cluster photos which are geographically close to each others. In this paper, we employ Kruskal's algorithm to cluster similar photos based on their geotag information. Let's assume a user $u$ visits at least $k$ different places (e.g. $k$ different cities) during the time interval $T$. We simply need to find the $k$ different clusters which represent the $k$ geographically different locations on the Earth. By recalling the power-law distribution for photos' locations, we expect that the cluster with the highest density (i.e. maximum number of photos) represents the user's hometown with a high probability. Therefore, we estimate the latitude and longitude of a user $u$'s place of living by taking a simple average over locations of all $u$'s photos' which fall inside the cluster with the highest density. Figures \ref{fig2} and \ref{fig3} demonstrate the locations of photos which have been taken by four different people (shown with red diamonds), their reported hometowns (shown with black circles), and the estimated hometowns by our algorithm (shown by green squares). As we can see in these figures, the photos can have a very wide geographical distribution. For instance, two users osiatynska and craignos have taken their photos from three different continents.

\begin{figure}
     \centering
     \includegraphics[width=3.00in, height=2.0in]{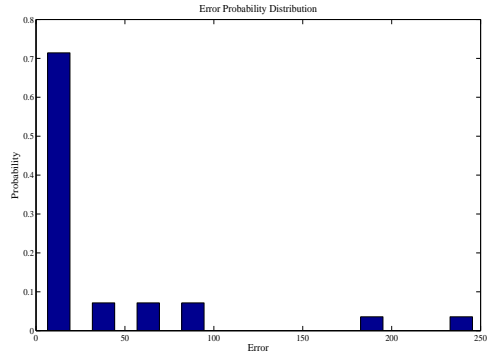}
     \caption{Error probability distribution}
     \label{fig4}
 \end{figure}

As mentioned earlier, we have estimated the possible hometowns for $31$ different people. To show the performance of our predictor, we can compute the distribution of distance error for predicted locations. Figure \ref{fig4} shows the probability distribution of errors for our predictor algorithm for these $31$ people. In $70\%$ of the cases our algorithm has predicted the place of living of people with low error.
 
\section{Conclusion}
This work shows the performance of using the geotagged contents for predicting people's places of living. Although our results might seem natural as we experience in our daily life, it highlights the importance of social profiles in predicting a photo's location. In other words, it proves the power of using social profiles for estimating the location of photos. %

\bibliographystyle{IEEEtran}
\bibliography{Ref}

\begin{thebibliography}{1}
\providecommand{\url}[1]{#1}
\csname url@samestyle\endcsname
\providecommand{\newblock}{\relax}
\providecommand{\bibinfo}[2]{#2}
\providecommand{\BIBentrySTDinterwordspacing}{\spaceskip=0pt\relax}
\providecommand{\BIBentryALTinterwordstretchfactor}{4}
\providecommand{\BIBentryALTinterwordspacing}{\spaceskip=\fontdimen2\font plus
\BIBentryALTinterwordstretchfactor\fontdimen3\font minus
  \fontdimen4\font\relax}
\providecommand{\BIBforeignlanguage}[2]{{%
\expandafter\ifx\csname l@#1\endcsname\relax
\typeout{** WARNING: IEEEtran.bst: No hyphenation pattern has been}%
\typeout{** loaded for the language `#1'. Using the pattern for}%
\typeout{** the default language instead.}%
\else
\language=\csname l@#1\endcsname
\fi
#2}}
\providecommand{\BIBdecl}{\relax}
\BIBdecl

\bibitem{Hays:im2gps}
J.~Hays and A.~A. Efros, ``im2gps: estimating geographic information from a
  single image,'' in \emph{Proceedings of the {IEEE} Conf. on Computer Vision
  and Pattern Recognition ({CVPR})}, 2008.

\bibitem{Kalogerakis:images2gps}
E.~Kalogerakis, O.~Vesselova, J.~Hays, A.~A. Efros, and A.~Hertzmann, ``Image
  sequence geolocation with human travel priors,'' in \emph{Proceedings of the
  {IEEE} International Conference on Computer Vision ({ICCV '09})}, 2009.

\bibitem{kleinsmall}
J.~Kleinberg, ``The small-world phenomenon: an algorithm perspective,'' in
  \emph{STOC '00: Proceedings of the thirty-second annual ACM symposium on
  Theory of computing}.\hskip 1em plus 0.5em minus 0.4em\relax New York, NY,
  USA: ACM, 2000, pp. 163--170.

\bibitem{nowell}
L.~D. Nowell, J.~Novak, R.~Kumar, P.~Raghavan, and A.~Tomkins, ``{Geographic
  routing in social networks},'' \emph{Proceedings of the National Academy of
  Sciences}, vol. 102, no.~33, pp. 11\,623--11\,628, 2005.

\bibitem{Backstrom:facebook}
L.~Backstrom, E.~Sun, and C.~Marlow, ``Find me if you can: improving
  geographical prediction with social and spatial proximity,'' in \emph{WWW
  '10: Proceedings of the 19th international conference on World wide
  web}.\hskip 1em plus 0.5em minus 0.4em\relax New York, NY, USA: ACM, 2010,
  pp. 61--70.

\bibitem{Gonzalez:nature}
\BIBentryALTinterwordspacing
M.~C. Gonzalez, C.~A. Hidalgo, and A.-L. Barabasi, ``Understanding individual
  human mobility patterns,'' \emph{Nature}, vol. 453, no. 7196, pp. 779--782,
  June 2008. [Online]. Available: \url{http://dx.doi.org/10.1038/nature06958}
\BIBentrySTDinterwordspacing

\end{thebibliography}

\end{document}